\documentclass[twocolumn,preprintnumbers,amsmath,amssymb,prl]{revtex4}

\usepackage{xspace}
\usepackage{graphicx}
\usepackage{dcolumn}
\usepackage{bm}
\usepackage{tabularx}
\usepackage{amsmath}
\usepackage{color}
\usepackage{ulem}
\DeclareUnicodeCharacter{2062}{}

\begin{document}

\title{Direct probe of magnetic field effects on phonons by ultrasound propagation in a quasi-two-dimensional honeycomb magnet Na$_2$Co$_2$TeO$_6$}

\author{Xiaochen Hong$^{1,2,3,}\footnote{hongxc@cqu.edu.cn}$, Maximilian Schiffer$^2$, Beat Valentin Schwarze$^4$, Marc Uhlarz$^4$, Xianghong Jin$^5$, Weiliang Yao$^5$, Lukas Janssen$^6$, Sergei Zherlitsyn$^4$, Bernd B\"{u}chner$^{3,7}$, Yuan Li$^{5,8}$, Young Sun$^{1}$ and Christian Hess$^{2,3,}\footnote{c.hess@uni-wuppertal.de}$}

\affiliation{
$^1$Department of Applied Physics and Center of Quantum Materials and Devices, Chongqing University, 401331 Chongqing, China\\
$^2$Fakult$\ddot{a}$t f\"{u}r Mathematik und Naturwissenschaften, Bergische Universit$\ddot{a}$t Wuppertal, 42097 Wuppertal, Germany\\
$^3$Leibniz-Institute for Solid State and Materials Research (IFW-Dresden), 01069 Dresden, Germany\\
$^4$Hochfeld-Magnetlabor Dresden (HLD-EMFL) and W\"{u}rzburg-Dresden Cluster of Excellence $ct.qmat$, Helmholtz-Zentrum Dresden-Rossendorf, 01328 Dresden, Germany\\
$^5$International Center for Quantum Materials, School of Physics, Peking University, 100871 Beijing, China\\
$^6$Institut f\"{u}r Theoretische Physik and W\"{u}rzburg-Dresden Cluster of Excellence $ct.qmat$, Technische Universit$\ddot{a}$t Dresden, 01062 Dresden, Germany\\
$^7$Institute of Solid State and Materials Physics and W\"{u}rzburg-Dresden Cluster of Excellence $ct.qmat$, Technische Universit$\ddot{a}$t Dresden, 01062 Dresden, Germany\\
$^8$Beijing National Laboratory for Condensed Matter Physics,
 Institute of Physics, Chinese Academy of Sciences, 100190 Beijing , China}

\date{\today}

\begin{abstract}
We study the phonon behavior of a Co-based honeycomb frustrated magnet Na$_2$Co$_2$TeO$_6$ under magnetic field applied perpendicular to the honeycomb plane. The temperature and field dependence of the sound velocity and sound attenuation unveil prominent spin-lattice coupling in this material, promoting ultrasound as a sensitive probe for magnetic properties. 
An out-of-plane ferrimagnetic order is determined below the Néel temperature $T_N=27$~K. 
A comprehensive analysis of our data further supports a triple-Q ground state of Na$_2$Co$_2$TeO$_6$.
Furthermore, the ultrasound data were systematically compared to the thermal transport results from literature, to unveil the importance of phononic contribution to the observed transport behaviors.
\end{abstract}

\pacs{not needed}

\maketitle

\section{Introduction}

Phonons, the quanta of lattice vibrations, are ubiquitous in solid-state systems.
Beyond their classical manifestations in thermodynamics, transport properties, and structural stability, phonons can couple strongly to electronic and magnetic degrees of freedom through elastoelectric and elastomagnetic interactions, giving rise to a rich array of emergent phenomena, including conventional superconductivity.
For quantum spin liquids (QSLs), most of which are good insulators, phonons can provide a unique window into their exotic quasiparticles by coupling to emergent magnetic excitations such as spinons, Majorana fermions, and emergent gauge fields \cite{Savary2017,Ye2020,Oh2025}.

However, the pervasiveness of phonons also poses significant challenges for reliably extracting tantalizing unconventional quasiparticle excitations.
For instance, thermal conductivity ($\kappa$) at low temperatures, a powerful probe for charge neutral excitations, is often limited by the difficulty of separating the putative unconventional contributions from the phonon background.
In practice, one needs either comparison with nonmagnetic reference compounds or performing a power-law fitting of the phononic $\kappa_{ph}(T)$ curve over a restricted temperature range. Both approaches are tricky and proliferate the discrepancies in interpreting the data.

This issue is exemplified in studies of $\alpha-$RuCl$_3$, the most prominent Kitaev QSL candidate.
Several key observations including complex magnetic field effect on $\kappa$ at low temperatures \cite{Leahy2017}, thermal Hall ($\kappa_{xy}$) effects \cite{Kasahara2018}, and quantum oscillations in $\kappa(H)$ \cite{Czajka2021}, were initially recognized as signatures of some QSL states.
Subsequent studies challenged them by trying to claim that magnetically perturbed phonons can also produce these novel phenomena \cite{Hentrich2018,Lefrancois2022,Bruin2022}. This controversy further intensifies when Na$_2$Co$_2$TeO$_6$ (NCTO) was found to exhibit some key experimental results similar to $\alpha-$RuCl$_3$ \cite{Yao2020, Lin2021, Hong2021, Sanders2022, Yang2022, Gillig2023, Chen2024, Zhou2024}. While NCTO initially emerged as another potential Kitaev QSL material \cite{Liu2018, Sano2018, Liu2020}, detailed theoretical calculations indicate that the Kitaev interaction, while present, is not the dominant term in NCTO’s Hamiltonian \cite{Winter2022}. Taken together, that implies there might be nothing beyond trivial contributions to $\kappa$ in NCTO, and that conclusion might also be true for $\alpha-$RuCl$_3$.
In this context, it is important to get more detailed knowledge of the phonons in this material class.

Ultrasound measurements provide a powerful experimental tool for directly probing phonon behavior in solids \cite{Luthi2005}. By analyzing changes in the phase and amplitude of ultrasound waves after propagation through a sample, and by monitoring their dependence on temperature and magnetic field, this technique enables high-resolution detection of variations in the sound velocity $\Delta v/v$ and sound attenuation $\Delta \alpha$ \cite{Zherlitsyn2014}.
This method has been extensively employed to study the phase diagrams of frustrated magnets \cite{Wosnitza2016,Zherlitsyn2014}, which often host a plethora of magnetic transitions due to competing interactions.
In particular, its application to spinel systems has demonstrated its effectiveness as a sensitive probe, capable of revealing subtle changes in lattice and spin dynamics \cite{Tuskan2021}.
Recently, ultrasound measurements have also been implemented to explore the elusive Majorana fermion under an in-plane magnetic field in $\alpha-$RuCl$_3$ \cite{Hauspurg2024,Hauspurg2025}.

In this work, we focus on measuring the impact of out-of-plane ($\bf B$//$\bf c$) magnetic fields on the in-plane ultrasound properties of NCTO. Our results demonstrate that ultrasound serves as a sensitive probe of magnetic transitions in NCTO. Very rich phenomena are revealed below its Néel temperature ($T_N=27$~K).
By analyzing our results, we confirm the existence of an out-of-plane ferrimagnetic order in NCTO and provide evidence supporting a triple-Q ground state \cite{Kruger2023,Chen2021}.
Furthermore, we explore the relationship between our ultrasound data and the reported thermal transport results under out-of-plane magnetic fields \cite{Yang2022,Hong2024,Li2023}, offering new insights into the interplay between lattice dynamics and magnetic excitations in this material and beyond.

\section{Experimental Method}

Single crystals of NCTO were prepared with a modified flux method \cite{Yao2020}. Although the typical thickness of the as-grown crystals are about 0.1~mm, very few of them can be thicker, preferred by an ultrasound experiment \cite{Yao2020}. Three of the thickest crystals, all with a thickness of 0.4$\pm$0.05~mm, were selected from different batches for this work.
NCTO has a hexagonal crystal structure (space group $P$6$_3$22) \cite{Zhang2023}.
The ultrasound propagation direction with the wave vector $k$ was set along the crystallographic zigzag $\bf a$-direction  (perpendicular to a Co-Co bond in the honeycomb plane, see the insets of Fig.~1) for simplicity. 
The polarization of the acoustic wave ($u$) was along $\bf a$, $\bf b$ or $\bf c$ crystallographic directions depending on the studied acoustic mode. The direction $\bf b$ is along Co-Co bond within the honeycomb plane.
The surfaces to which the transducers would be attached were carefully polished against ultra-fine grade abrasive papers.

Ultrasound measurements were performed using a pulse-echo phase-sensitive detection technique \cite{Zherlitsyn2014}. To generate and detect ultrasonic signals in the frequency range 20–150~MHz, a pair of LiNbO$_3$ ultrasound transducers (36$^{\circ}$ Y-cut for the longitudinal and 41$^{\circ}$ X-cut for the transverse waves) were attached to both sides of the sample with a liquid polymer Thiokol LP-32.
The experiments were carried out in a physical property measurement system (PPMS) from Quantum Design and a dilution refrigerator from Oxford Instruments, both equipped with a superconducting magnet.

\section{Results and Discussion}

\begin{figure}
\includegraphics[clip,width=0.37\textwidth]{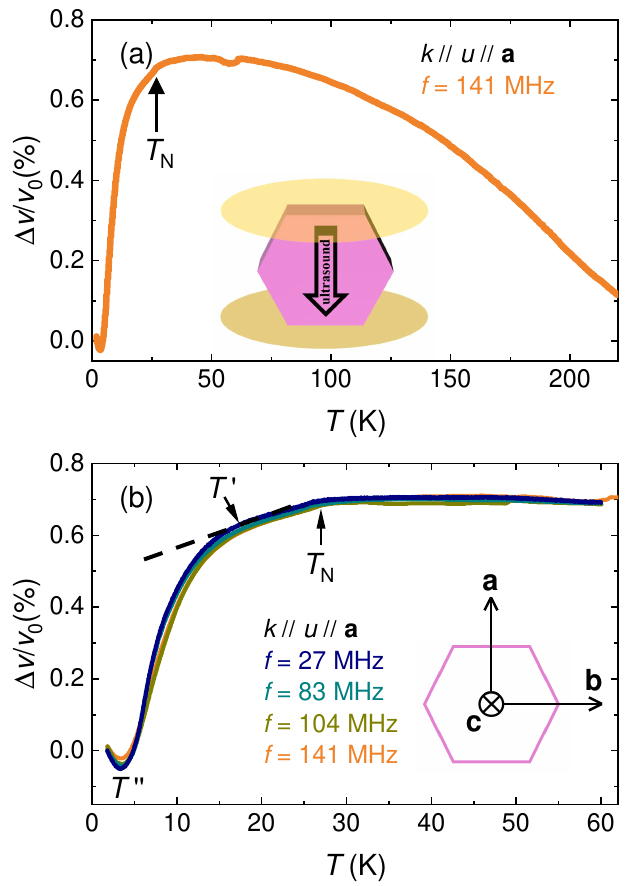}
\caption{
(a) Representative temperature dependence of $\Delta v/v_0$, collected in the LA mode at the ultrasound frequency $f\approx141$~MHz. (b) The low temperature data at different ultrasound frequencies. Three anomalies $T_N=27$~K, $T'=16$~K, and $T''=4$~K can be addressed. The cartoons illustrate the experimental geometry.}
\end{figure}

The temperature dependence of the relative changes in the sound velocity $\Delta v/v_0$ for the longitudinal acoustic (LA) mode is shown in Fig.~1(a). 
The overall $\Delta v/v_0(T)$ curve reveals the lattice hardening upon cooling, and a softening appears below the magnetic transition temperature $T_N$.
The data at low temperature is enlarged in Fig.~1(b), plotted together with results taken with various ultrasound frequencies, to check for a possible frequency dependence.
All these curves overlap almost perfectly. 
Three anomalies can be reliability resolved: the primary antiferromagnetic transition at $T_N=27$~K and two additional features at $T'=16$~K and $T''=4$~K. 
Although this hierarchy of magnetic transitions below $T_N$ in NCTO has been previously documented \cite{Lefrancois2016,Bera2017}, it remains scarcely addressed.
A recent comprehensive study combined spectroscopic and scattering techniques attributed this behavior to the interplay between two-dimensional and three-dimensional magnetic orders in NCTO \cite{Chen2021}.

A set of field-dependent $\Delta v/v_0$ isotherms are presented in Fig. 2(a). Above 5~K, the $\Delta v/v_0(H)$ curves show moderate lattice softening in field without additional features up to $7.5$~T. The amplitude of changes in $\Delta v/v_0(H)$ increases considerably upon cooling.
A clear dip in $\Delta v/v_0(H)$ can be resolved at 3~K.
This dip gradually shifts to a lower field at lower temperatures.
At the lowest temperature of 32~mK another broader dip at round 16~T was detected.
These features are summarized in Fig.~2(b) in the phase diagram of NCTO for an out-of-plane magnetic field, together with the transitions observed in the magnetic susceptibility \cite{Zhang2024b}.
While all the ultrasound anomalies can find their correspondence in magnetization, some magnetic transitions reported in the literature leave no detectable fingerprint in our ultrasound data, including $H_1$ and $H_0$ above 3~K ~\cite{Zhang2024b,Bera2023}.
This fact means that the acoustic modes in NCTO are selectively coupled to the spins.
For the studied acoustic modes, the magnetoelastic coupling can be very small at some phase transitions in NCTO.

\begin{figure}
\includegraphics[clip,width=0.49\textwidth]{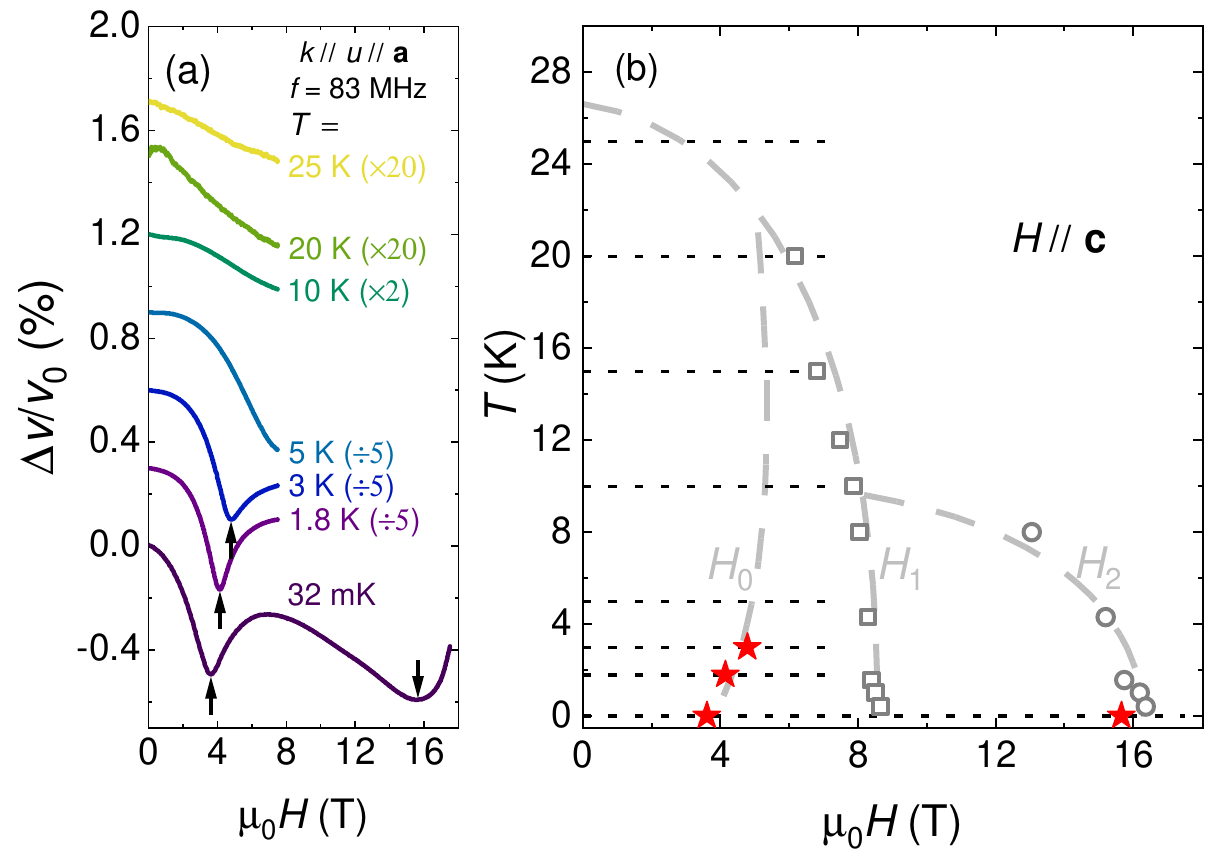}
\caption{
(a) Field dependence of $\Delta v/v_0$ at different temperatures. The curves are shifted vertically for clarity. Note the different rescale factors for each curve. The black arrows indicate the dips in some curves.
(b) The gray circles and squares are phase boundaries of NCTO inferred from its magnetic susceptibility data \cite{Zhang2024b}. The red stars are the positions of anomalies from Fig.~2(a). We follow the notation for the critical fields $H_1$ and $H_2$ from Ref. \cite{Zhang2024b}.
The dashed gray lines are guide-to-the eyes of the boundaries of the contour plot of $dM/dH$ data in Ref.~\cite{Zhang2024b}.
Although there is a clear line below $H_1$ observable, it was not labeled as a critical field in Ref.~\cite{Zhang2024b}. Nevertheless, an electron spin resonance (ESR) work identified it also as a phase transition \cite{Bera2023}. Thus we use $H_0$ to name this line. The dashed black lines indicate the parameter space reached in Fig.~2(a).
}
\end{figure}

\begin{figure}
\includegraphics[clip,width=0.3\textwidth]{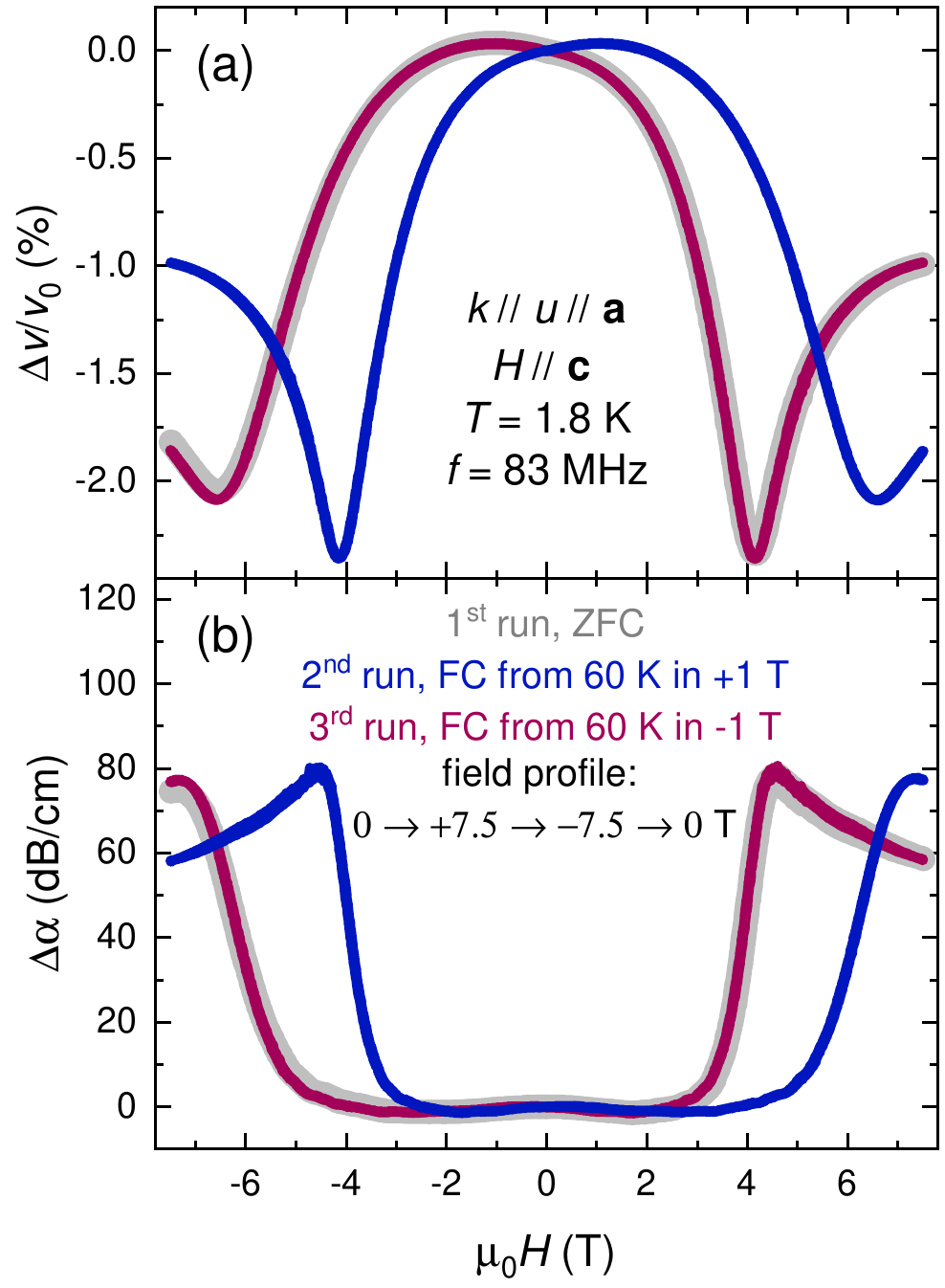}
\caption{Field dependence of (a) $\Delta v/v_0$ and (b) $\Delta \alpha$ at 1.8~K, measured after cooled in opposite field directions. The first field-sweep run was done after the sample was cooled in zero, or in small ($<14$~mT) remanent field of our magnet. The second and third measurements were done after cooling in a finite field applied along opposite directions.}
\end{figure}

Recently, there have been some debates on whether the magnetic ground state of NCTO is an unusual triple-Q state or the conventional multi-domain zigzag order \cite{Chen2021,Yao2023,Jin2025,Zhang2023,Dhanasekhar2025}.
To shed more light on this issue, we investigated field dependence of $\Delta v/v_0$ and $\Delta \alpha$ under different cooling protocols (Fig.~3).
We observe a marked field-asymmetry tied to cooling conditions, switchable by modifying the field-cooling procedure above $T_N$.
This provides clear evidence of an out-of-plane ferromagnetic moment, consistent with the reported out-of-plane ferrimagnetic behavior in NCTO \cite{Yao2020}.
There are two mechanisms proposed for the ferrimagnism in NCTO, both invoking the canting of the in-plane magnetic moments. Some previous work suggested a triple-Q ground state in NCTO and several other Co-based quasi-2D honeycomb magnets \cite{Chen2021,Yao2023,Jin2025,Kruger2023,Francini2024}. Another scenario insists that many cobaltates manifest a zigzag ground state stabilized within an XXZ model \cite{Zhang2023,Zhang2024b,Dhanasekhar2025,Winter2022}, and the out-of-plane ferrimagnetism is due to unbalanced domains.
Note that the zero-field-cooled (ZFC) $\Delta v/v_0(H)$ and $\Delta \alpha(H)$ data perfectly match the field-cooled (FC) data of the fully-switched states.
It is unlikely that the alignment of domains can be made by a small remanent field. Hence, a more natural explanation is provided by a domain-free triple-Q ground state of NCTO.

To gain more insight, one set of additional $\Delta v/v_0(H)$ results at elevated temperatures all the way up to above $T_N$ are shown in Fig.~4(a). Results collected with different ultrasound frequencies and polarizations can be found in Supplemental Material \cite{SM}.
Obviously, the $\Delta v/v_0(H)$ curves are weakly field-suppressed and quite symmetric at temperatures higher than $T_N$.
A butterfly-shaped hysteresis starts to grow at low temperatures. The hysteretic region and asymmetry increase as the temperature is lowered. Finally, the hysteresis quickly shrinks, leaving very asymmetric $\Delta v/v_0(H)$ curves below about 12~K.

\begin{figure}
\includegraphics[clip,width=0.49\textwidth]{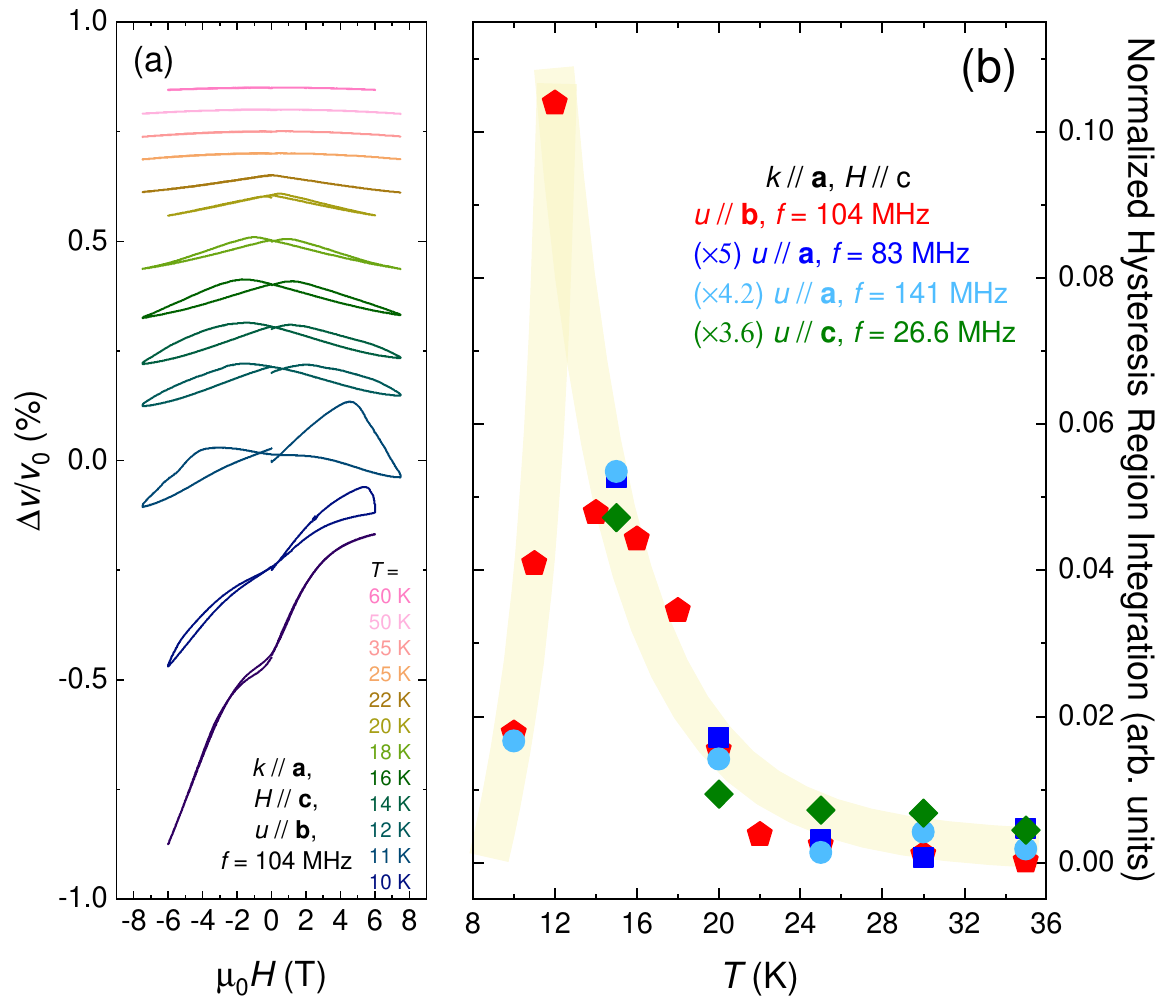}
\caption{(a) $\Delta v/v_0(H)$ isotherms measured in NCTO in opposite field directions. The curves are shifted vertically for clarity. 
(b) The temperature dependence of the integrated hysteretic region of $\Delta v/v_0(H)$ isotherms normalized to the maximum change of $\Delta v/v_0$ at each temperature.
The integration of $\Delta v/v_0(H)$ hysteresis for data collected under different experimental settings (see text) are rescaled to show a common temperature behavior. The yellow stripe is a guild to the eye.
}
\end{figure}

\begin{figure*}
\includegraphics[clip,width=0.9\textwidth]{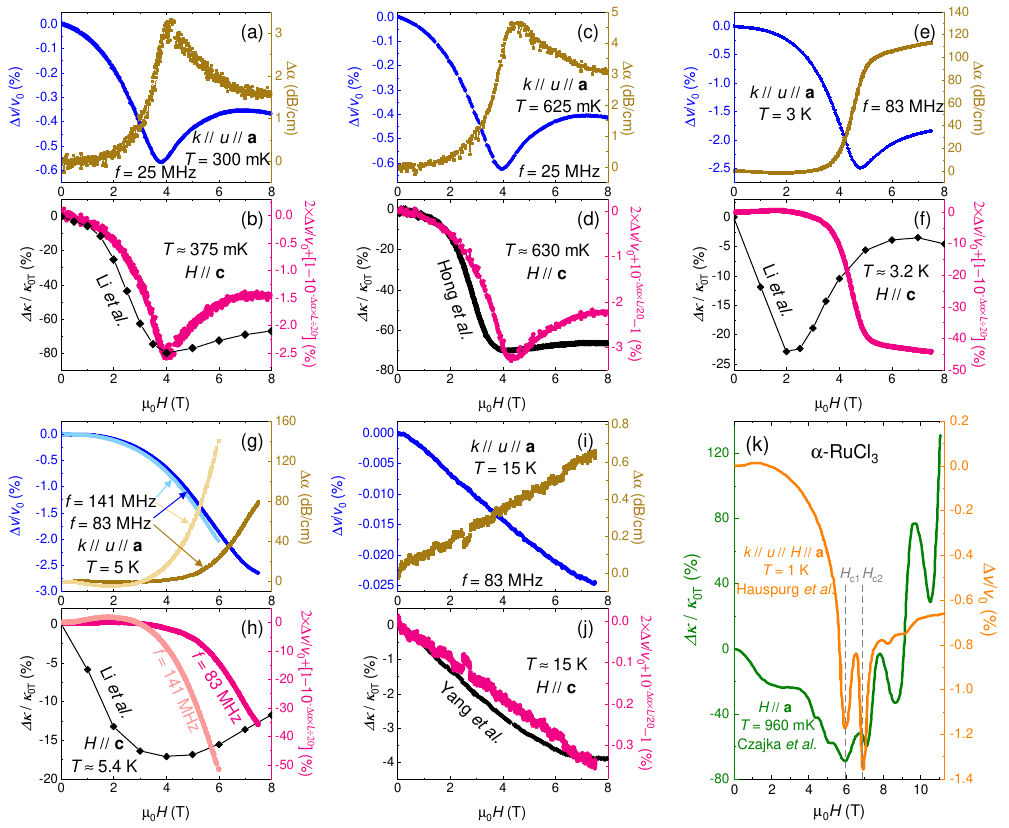}
\caption{Field dependence of $\Delta v/v_0$ and $\Delta \alpha$ at selected temperatures: (a) 300~mK, (c) 625~mK, (e) 3~K, (g) 5~K, and (i) 15~K. 
The corresponding estimates of the change of phonon conductivity (see text) are plotted in red in (b), (d), (f), (h) and (j), respectively. Field dependence of the thermal conductivity $\Delta\kappa/\kappa_{0T}$ at similar temperatures reported in the literature is also shown \cite{Yang2022,Hong2024,Li2023}. For $T=$5~K, the ultrasound data of two frequencies were presented. 
(k) Field dependence of $\Delta\kappa/\kappa_{0T}$ and $\Delta v/v_0$ of $\alpha-$RuCl$_3$ at comparable temperatures, with the field applied along the same in-plane crystallographic direction. Data are from Refs.~\cite{Hauspurg2024,Czajka2021}.}
\end{figure*}

We use the integral of the enclosed region of the $\Delta v/v_0(H)$ curves to quantify the degree of hysteresis. 
For each $\Delta v/v_0(H)$ curve displaying hysteresis, the integral over the positive field branch (between 0~T and 7.5~T) differs from that over the negative branch (between 0~T and -7.5~T), with the larger integral determined by the cooling and field history.
In Fig.~4(b), the larger integral was taken for each curve and normalized to the full amplitude of $\Delta v/v_0$ within the corresponding measurement batch, i.e., [$\int_{0}^{H}(\Delta v_{top}-\Delta v_{down})/v_0$]/$[\Delta v/v_0(H)]|_{max}$.
While ultrasound frequency has little impact on $\Delta v/v_0(H)$ [e.g., Fig.~1(b)], the acoustic wave polarization distinctly alters the amplitude and may also affect the line-shape \cite{SM}.
Thus, data acquired under different experimental settings were rescaled to enable direct comparison in the same figure, revealing a common temperature-dependent behavior for different acoustic modes.
The different rescale factors for the two $u$//$\bf a$ datasets respects to their different field range (up to 6~T and 7.5~T, respectively) \cite{SM}.
Notably, the hysteresis grows rapidly below $T_N$, reaches a maximum at around 12~K, and then drops quickly toward zero at lower temperature.

Naively, one might anticipate stronger field hysteresis corresponding to more substantial reorganization of spin configurations.
If NCTO has a XXZ ground state with domains developing below $T_N$, the spin-flip energy barrier should exhibit monotonic enhancement with decreasing temperature. Consequently, the observed hysteresis maximum occurring well below $T_N$ is an unconventional behavior.
On the other hand, a recent Monte Carlo simulation work deduced a compensation point $T_{cp}$ at roughly $T_N/2$ for a sign reversion of the residual magnetization out of the honeycomb plane, if a triple-Q model is adopted for NCTO \cite{Francini2024}.
Given that the energy difference between two magnetic configurations vanishes near 12~K, an applied magnetic field within this temperature regime becomes particularly effective in modifying the magnetic order.
Taken together, our observations are better interpreted by a triple-Q ground state of NCTO.

Finally, the effect of out-of-plane field on the thermal conductivity $\kappa(H)$ of NCTO are scrutinized with respect to the phonon properties. 
It is known that phonons dominate thermal transport in NCTO \cite{Hong2024}. Since an out-of-plane magnetic field affects both the phonon speed and scattering rate, it may have dual effects on $\kappa$:
\begin{equation}
\kappa \propto Cvl=Cv^2\tau \propto \frac{Cv^2}{\Gamma}~~,
\end{equation}
where $l$ is the mean free path and $\tau$ is the relaxation time of phonons, which is inversely proportional to the scattering rate $\Gamma$. The phonon specific heat $C$ is (to first order) field independent. 
The field-induced change of $\kappa$ can be written as
\begin{equation}
\Delta \kappa \propto \frac{C(v_0+\Delta v)^2}{\Gamma_0+\Delta\Gamma}-\frac{Cv_0^2}{\Gamma_0}~~.
\end{equation}
If considered to the lowest order, this can be approximated 
\begin{equation}
\frac{\Delta \kappa}{\kappa} \approx 2\times\frac{\Delta v}{v_0}-\frac{\Delta\Gamma}{\Gamma_0+\Delta\Gamma}~~,
\end{equation}
with the first and second term manifesting the contributions dominated by the field change of phonon speed and scattering rate, respectively.
For pulse-echo ultrasound experiments, the damping term $\Delta \alpha$ is defined as 
\begin{equation}
\Delta \alpha \equiv -\frac{20}{L} \log_{10}\frac{I}{I_0}~~,
\end{equation}
where $L$ is the effective sample length, and $I$ is the ultrasound-signal amplitude \cite{Zherlitsyn2014,Hauspurg2024}.
By assuming that the scattering of phonons reduces the ultrasound signal and thermal transport in the same way, we get
\begin{equation}
\frac{\Gamma_0}{\Gamma} = \frac{\tau}{\tau_0} \propto \frac{I}{I_0}=10^{-\Delta \alpha \times L/20}~~.
\end{equation}
Substituting Eq.~(5) into Eq.~(3), we can write
\begin{equation}
\frac{\Delta \kappa}{\kappa} \approx 2\times\frac{\Delta v}{v_0}+10^{-\Delta \alpha \times L/20}-1~~.
\end{equation}

The estimated relative change in phonon thermal conductivity based on ultrasound data is shown in Fig.~5, by taking the sample thickness $L=0.4$ mm in Eq.~(6). The effect of the out-of-plane magnetic field on the thermal transport of NCTO at nearly the same temperature reported in the literature is also sketched for comparison \cite{Yang2022,Hong2024,Li2023}.
At temperatures below 1~K [Fig.~5(b) and (d)] and at around 15~K [Fig.~5(j)], the shape of the calculated and measured $\Delta\kappa(H)$ curves roughly matches, but the amplitude calculated from the phonon data is underestimated by an order of magnitude or more. The reason might be that only one acoustic mode at a single frequency is considered in this work, while all phonon modes contribute to phonon thermal transport. Actually, the $\Delta \alpha$ term is strongly frequency dependent \cite{SM}.
It is justified to claim that these results are consistently explained by the dominant phonon contribution of total thermal conductivity.
However, as clearly demonstrated by Fig. 5(f) and (h), at intermediate low temperatures of 3~K and 5~K, while an impressively large field suppression of $\kappa$ is expected from the ultrasound data, the measured $\Delta\kappa(H)$ are relatively modest (overestimated by a factor of 2), and its shape does not follow the phonon expectation.
Hybridization of phonons and magnons was proposed by Ref. \cite{Li2023} to explain the thermal Hall experimental results observed in NCTO. We argue that such hybridization might also be responsible for the discrepancy between real $\Delta\kappa(H)$ and the estimation from our ultrasound data in the intermediate low temperature range. If phonons and magnons exhibit synergistic transport behavior, the simplified scattering picture underlying Eq. 6 becomes inadequate. We expect that the comparison between ultrasound and thermal transport data can help delineate the temperature range where the hybridization effect prevails.

The literature $\Delta v/v_0(H)$ and $\Delta\kappa(H)$ curves of $\alpha-$RuCl$_3$ at the similar temperature are plotted in Fig.~5(k) \cite{Hauspurg2024,Czajka2021}. 
Both show clear oscillatory structures above the two well-defined critical fields. 
This oscillatory structure in $\kappa(H)$ was initially attributed to some charge neutral magnetic excitations \cite{Czajka2021}, and was later challenged by a more conventional scenario of strongly scattered phonon transport channels \cite{Bruin2022,Lefrancois2023}.
In the context of our results and discussion, the scattered phonon explanation and possibility of phonon-magnon hybridization should also be carefully considered when interpreting these novel $\kappa(H)$ data.

\section{Summary}

To summarize, using the ultrasound technique we studied the temperature and field induced transitions in NCTO.
We observed strongly asymmetric and hysteretic features in the acoustic properties of NCTO in the magnetic field applied perpendicular to the honeycomb plane in opposite directions.
Quantitative analyses of the hysteresis area unveiled a sharp maximum reached at around 12~K, with a strong suppression when the temperature is further lowered. 
This observation suggests the scenario with triple-Q ground state in NCTO rather than a zigzag ground state proposed from an XXZ model.
A comparative analysis of our ultrasound measurements with previously reported thermal transport data reveals the pivotal role of phonons in governing the complex thermal transport properties of NCTO at low temperatures \cite{Yang2022,Hong2024,Li2023}. These findings underscore the necessity of carefully accounting for phonon-related effects when interpreting experimental results in frustrated magnetic systems.

\section{Acknowledgments}
We thank J\'{e}r\'{e}my Sourd and Andreas Hauspurg for technical support.
We thank Robin Neumann for valuable feedback on the manuscript and for sharing their experimental data with us.
This work has been supported by the Deutsche Forschungsgemeinschaft (DFG) through Project No. 247310070 (SFB 1143), Project No. 390858490 (W\"{u}rzburg-Dresden Cluster of Excellence {\it ct.qmat}, EXC 2147), and Project No. 411750675 (Emmy Noether program, JA2306/4-1).
This project has received funding from the European Research Council (ERC) under the European Unions-Horizon 2020 research and innovation programme (grant agreement No. 647276-MARS-ERC-2014-CoG).
We acknowledge support of the HLD at HZDR, member of the European Magnetic Field Laboratory (EMFL).
XCH acknowledge the start-up funding of Chongqing University.

\section{DATA AVAILABILITY}
The data that support the findings of this article are openly available \cite{Data}.

\end{document}